\documentclass[10pt,pra,aps,showpacs,twocolumn,superscriptaddress]{revtex4-1}
\usepackage{graphicx,bm}
\usepackage{amsmath}
\usepackage{amssymb}
\usepackage{epsfig}
\usepackage{epsf}
\usepackage{verbatim}
\usepackage{hyperref}
\usepackage[usenames]{color}

\newcommand{\bs}{\boldsymbol}

\def\be#1\ee{\begin{equation}#1\end{equation}}
\def\ba#1\ea{\begin{align}#1\end{align}}

\begin{document}

\title{
Spin Hall mode in a trapped thermal Rashba gas
}

\author{J. Armaitis}
\email{jogundas.armaitis@tfai.vu.lt}

\author{J. Ruseckas}
\affiliation{Institute of Theoretical Physics and Astronomy,
Vilnius University,
Saul\.{e}tekio Ave.\ 3, LT-10222 Vilnius, Lithuania}

\author{H. T. C. Stoof}
\affiliation{Institute for Theoretical Physics
Utrecht University,
Princetonplein 5, 3584 CC Utrecht, The Netherlands}
\affiliation{Center for Extreme Matter and Emergent Phenomena,
Utrecht University,
Princetonplein 5, 3584 CC Utrecht, The Netherlands}

\author{R. A. Duine}
\affiliation{Institute for Theoretical Physics
Utrecht University,
Princetonplein 5, 3584 CC Utrecht, The Netherlands}
\affiliation{Center for Extreme Matter and Emergent Phenomena,
Utrecht University,
Princetonplein 5, 3584 CC Utrecht, The Netherlands}
\affiliation{Department of Applied Physics, Eindhoven University of
Technology, P.O. Box 513, 5600 MB Eindhoven, The Netherlands}

\date{\today}
\begin{abstract}
We theoretically investigate a two-dimensional harmonically-trapped gas
of identical atoms with Rashba 
spin-orbit coupling and no interatomic interactions.
In analogy with the spin Hall effect in uniform space, the gas
exhibits a spin Hall mode. 
In particular, in response to a displacement of the
center-of-mass of the system, spin-dipole moment oscillations occur.
We determine the properties of these oscillations exactly, and find that their
amplitude strongly depends on the spin-orbit coupling
strength and the quantum statistics of the particles.
\end{abstract}

\maketitle

\section{Introduction}

Collective behavior provides an important pathway towards the measurement of
various physical properties in a multitude of systems. In particular,
material properties pertaining to transport often leave an imprint in the
collective motion of the system. For example, the charge carrier density can be
determined by observing the Hall response \cite{Preston1991}. Various
magnetic oscillations allow to measure the effective mass of carriers, as
well as to quantify the level of disorder in the material \cite{Shoenberg1984}. 
The speed of sound in a material
allows to access information about such mechanical properties as shear modulus,
density, and compressibility \cite{Rossing2014}. Recently, vorticity has been 
identified as a hallmark of viscous electron transport in graphene 
\cite{Levitov2016, Bandurin2016}.

When it comes to ultracold atomic gas systems, collective modes have been at
the center of the field since the very first experiments \cite{Andersen2004,
Giorgini2008}. 
By ``collective modes'' in this context it is meant that the
system is perturbed as a whole. The response to this
perturbation is typically deduced from measurements that probe
the full system, and not its constituents, even in the absence of interparticle interactions.
Arguably the simplest of such modes is the center-of-mass
oscillation (also known as the dipole mode) of the 
whole cloud of atoms in a harmonic trap. 
Kohn's theorem states that this mode does not decay, is not
affected by interactions, and that its frequency equals that of the trap
\cite{Kohn1961, Bamler2015}. 
This theorem does not apply in the presence of
spin-orbit coupling (SOC). SOC breaks Galilean invariance
\cite{Dalibard2011, Goldman2014, Stringari2016, Stringari2017},
as demonstrated by the altered frequency of the dipole mode
\cite{Zhang2012b, Chen2012}.
Another case when Kohn's theorem does not apply is
an out-of-phase oscillation of two interacting 
species of atoms in the same trap, i.e., a so-called
(pseudo)spin-dipole oscillation. 
To probe this mode, a species-dependent force displaces the clouds of the two
species with respect to each other while keeping the center of mass of the
whole system at the bottom of the trap.   
Such oscillations have been observed in various experiments
with bosons \cite{Hall1998,Modugno2002,Koller2015,Bienaime2016}, 
fermions \cite{DeMarco2002, Valtolina2016}, as well as mixtures of bosons
and fermions \cite{Ferrier-Barbut2014, Delehaye2015, Roy2017}.

\begin{figure}[t]
\begin{center}
\includegraphics[width=0.75\linewidth]{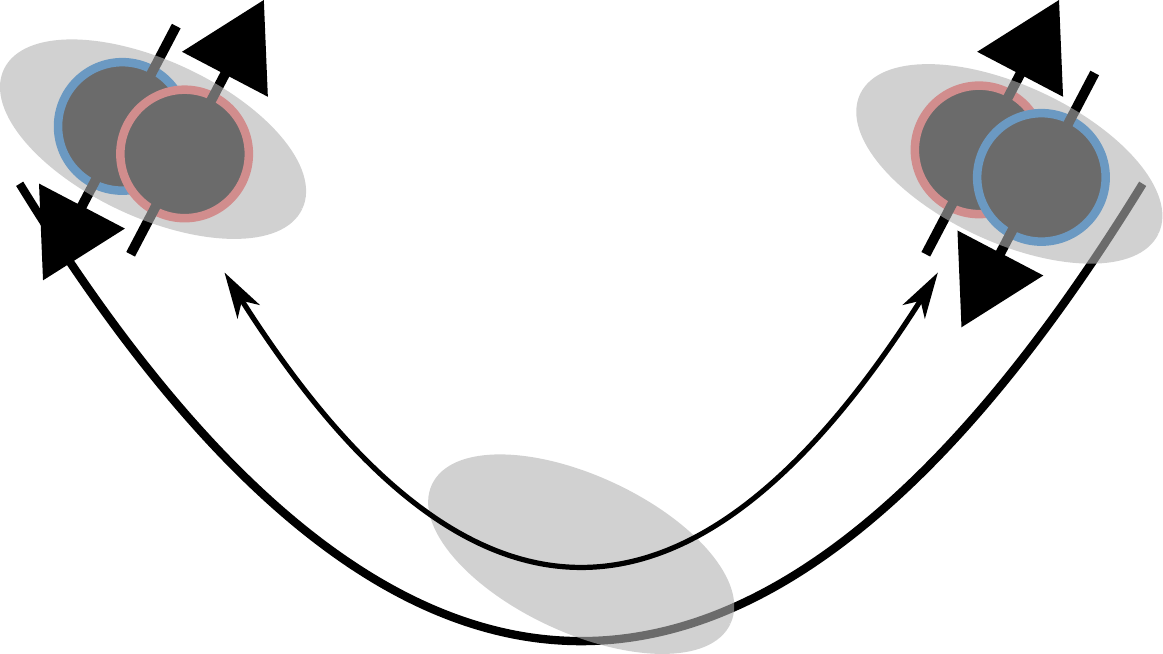}
\caption{
Spin Hall mode: due to the spin Hall effect, a Rashba 
spin-orbit-coupled system (grey ellipsoid) in a harmonic trap (thick parabola) 
exhibits oscillations of the transverse spin-dipole moment in response to a
displacement away from the bottom of the trap. During these oscillations the
whole cloud swings back and forth along the axis of displacement, as indicated 
by the thin arrow. More importantly, the transverse spin-dipole moment also
oscillates, whereas the total spin of the cloud remains zero. 
This is visualized in the figure by the alternating separation of the
average positions of the spin-up (red) and the spin-down (blue)
particles. The spin quantization axis is
perpendicular to the spin-orbit-coupling plane, whereas the spin-separation 
axis is perpendicular to both the spin axis, and the direction of 
displacement.
\vspace{-8mm}}
\label{fig:exp0}
\end{center}
\end{figure}

A similar excitation in response to a spin-independent force
known as the intrinsic spin Hall effect has been central to the field of spintronics
\cite{Sinova2004, Sinova2015}. There, a transverse spin current is generated
in response to a longitudinal charge current in a uniform system. 
This occurs generically due to the presence of a
(Rashba \cite{Manchon2015} or Dresselhaus \cite{Dresselhaus1955}) SOC,
since the spin of different momentum states precesses differently
in response to a spin-independent force.
Building up on this knowledge, we analyze such a situation in the
presence of a trapping potential, where one expects a similar
response to occur (Fig.~\ref{fig:exp0}).
Contrary to the spintronics nomenclature, in the ultracold-atom literature,
the spin Hall effect almost \cite{Armaitis2017a} always refers to the two
eigenstates of a system with one-dimensional (also known as equal 
Rashba-Dresselhaus) SOC experiencing opposite transverse forces in the 
absence of spin precession \cite{Beeler2013, Kennedy2013}. Spin
is correspondingly conserved in one direction in these ultracold-atom
experiments. However, this does not have to be the case in general.
Indeed, spin is typically not conserved
in solid-state spintronics experiments \cite{Sinova2015}.

Several different schemes of inducing a two-dimensional SOC in ultracold atom
systems have been proposed
\cite{Sau2011, Campbell2011, Anderson2013, Campbell2016} and at least one of
them has been experimentally realized \cite{Huang2016}. Hence, we are
motivated to investigate the collective modes of a harmonically confined gas
in the presence of a two-dimensional SOC. Furthermore, since all the
experimentally realistic two-dimensional SOC schemes involve some degree of
heating \cite{Dalibard2011, Goldman2014, Eckardt2017}, at least the early
experiments are likely to operate in the thermal-gas regime.  Therefore, we
study quantum degenerate thermal, i.e.\ noncondensed, fermions and bosons. 
In this regime energy scales set by various
scattering processes (interparticle interactions) are small compared to the
thermal energy, allowing us to neglect interaction effects in this study. Moreover,
we do not consider the constant Zeeman terms leading to an anomalous Hall effect
\cite{Bijl2011} here. This allows us to stay in the regime where the
transverse Hall response
only exists in the spin channel.

Our main finding is that a collective mode in the trap analogous
to the spin Hall response in the uniform system is indeed present.
We call this mode the spin Hall mode.
The amplitude of this response depends in a nontrivial
way on the SOC strength, and is different for Bose
and Fermi particles
(Figs.~\ref{fig:boseThermal}~and~\ref{fig:fermiThermal}).

This paper is organized as follows. In Sec.~\ref{sec:framework} we 
present the problem and introduce the notation. 
We treat the weak-SOC and 
strong-SOC limits analytically for the lowest-lying states
in Sec.~\ref{sec:analytic}.
In Sec.~\ref{sec:numerical}
we present an exact numerical treatment of
both bosonic and fermionic degenerate gases at nonzero temperature.
Finally, Sec.~\ref{sec:summary} summarizes our results and 
provides some directions for future work.

\section{Theoretical framework}
\label{sec:framework}

\subsection{System}

The Hamiltonian of our system
\be
H=H_K+H_{R}+H_{T}
\ee
consists of the kinetic energy term $H_K$, the Rashba SOC term $H_{R}$, and
the harmonic trapping potential $H_{T}$. 
It is assumed from the outset that a steep trapping potential in the $z$ 
direction
dominates all other energy scales, allowing
us to concentrate on the dynamics in the $x$--$y$ plane.
It is convenient
to treat the problem in the units where the reduced Planck constant,
particle mass and trap frequency are set to unity, $\hbar=M=\omega_T=1$.
In these units,
\ba
H_K = \bs p^2/2 = (p_x^2 + p_y^2)/2, \\
H_T = \bs x^2/2 = (x^2 + y^2)/2,
\ea
where vectors are denoted by the bold font, $\bs x$ is a vector of coordinate operators, and $\bs p$ is a vector of momentum operators.
We keep identity matrices in spin space implicit throughout the article.
The Rashba SOC Hamiltonian is
\be
H_R = v(\sigma_x p_y - \sigma_y p_x),
\ee
where $v$ is the Rashba SOC strength, and $\sigma_{i}$ are the
 Pauli matrices. Using
polar coordinates in the plane, $p_{x}=p\cos\theta$ and $p_{y}=p\sin\theta$,
as well as introducing the spin raising and lowering operators
\ba
\sigma_{x} &= \sigma_{+}+\sigma_{-}, \\
i\sigma_{y} &= \sigma_{+}-\sigma_{-},
\ea
the SOC part of the Hamiltonian can be written as
\be
H_{R}=ipv\left(\sigma_{+}e^{-i\theta}-\sigma_{-}e^{i\theta}\right).
\ee
The two eigenspinors of $H_{R}$, namely,
\be
\chi_\pm=\frac{1}{\sqrt{2}}
\begin{pmatrix}
\pm ie^{-i\theta}\\
1
\end{pmatrix}
\ee
only depend on the
direction of the momentum in the plane, and not on its magnitude.
In the absence of the trap, the energies of the two branches corresponding to these eigenspinors
are
\be
E_\pm=\frac{p^{2}}{2}\pm pv.
\label{eq:rashbaDisp}
\ee

The Hamiltonian $H$ and its various extensions have already been the subject 
of several investigations.
Most of the work has concentrated on the study of Bose-Einstein condensates and their
dynamics 
\cite{Larson2009,Sinha2011,Wang2010,Yip2011, Hu2012a, Hu2012b, Zhang2012a}.
Some attention has been paid to the interplay between interactions
and SOC \cite{Yu2011, Yin2014, Schillaci2015}, as well as rotations and SOC \cite{Doko2016,Doko2017}.
Moreover, various properties of this system in anisotropic (deformed) trapping potentials
have been investigated \cite{Marchukov2013,Marchukov2014,Marchukov2015}.
Regarding dynamics, it was shown that a sudden ramp up of the SOC strength
initiates collapse and revival dynamics of the total magnetization 
for a Fermi gas \cite{Natu2013}.

\subsection{Moments and modes}

In general, one may excite the system using an operator
$O_1$ and witness the response of the system to this perturbation
in the evolution of the expectation
values of an operator $O_2$. The response to excitations are
intimately related to the eigenmodes of the system.

In a trapped system of ultracold atoms,
a natural excitation is a small shift of the center of mass of the
system (or, equivalently, the bottom of the trapping potential). Such
a perturbation with an infinitesimal amplitude $|\bs x_0|\ll 1$ in a
direction $\hat{\bs x}_0 = \bs x_0/|\bs x_0|$ is described by a translation operator
\be
T = 1 - i \bs p \cdot \bs x_0.
\label{eq:genTranslation}
\ee
In general, this excitation may result both in longitudinal response 
(along $\hat{\bs x}_0$), 
and transverse response (along $\hat{\bs x}_0^\perp$, which is
perpendicular to $\hat{\bs x}_0$). 

Arguably the simplest observable is the center-of-mass 
position, namely, $\langle \bs x \rangle$.
In a trapped system in the absence of SOC, the response to $T$ 
is fully described by a single-frequency oscillation in the longitudinal
channel of the center of mass position of the cloud,
$\langle \bs x \cdot \hat{\bs x}_0 \rangle$,
and is known as the dipole mode or Kohn's mode as described
in the introduction.
In the presence of SOC this mode is
significantly modified \cite{Zhang2012b}, as will also be 
demonstrated in the subsequent discussion.
The center of mass response in the transverse channel,
$\langle \bs x \cdot \hat{\bs x}_0^\perp \rangle$,
corresponds to the anomalous Hall effect. It requires
a nonzero Berry curvature in momentum space, which
can for instance be achieved by adding a Zeeman
term to our SOC Hamiltonian. An oscillation of
the center-of-mass position transverse to the direction of the
excitation is therefore called the anomalous Hall mode \cite{Bijl2011}. 

Even though expectation values of various other operators are also
accessible to ultracold atom experiments (see, e.g., Ref.~\cite{StamperKurn2013}),
we limit our discussion to the spin-dipole moment 
$\langle x_i \sigma_j \rangle$. As mentioned earlier, it is possible
to initialize an ultracold-atom system in a state with a nonzero 
spin-dipole moment by separating out the two spin states in position space
\cite{Hall1998,Modugno2002,Bienaime2016, DeMarco2002, Valtolina2016, 
Ferrier-Barbut2014, Delehaye2015, Roy2017}. Subsequently, a weakly-interacting
system in the absense of SOC exhibits spin-dipole
oscillations due to harmonic confinement, known as the spin-dipole mode. 

From the perspective of spintronics, a transverse spin-dipole 
moment or spin accumulation
$\langle \hat{\bs x}^\perp_0 \sigma_z\rangle$, which emerges in response to a spin-independent
perturbation $T$ or voltage, is known as the spin Hall effect \cite{Sinova2004}. 
Here we consider exactly such a setup. Namely, we start from a state
with a vanishing spin-dipole moment,
apply a spin-independent perturbation, Eq.~\eqref{eq:genTranslation},
and subsequently observe oscillations in both the longitudinal center-of-mass position
and transverse spin-dipole moment. It is therefore natural
to call this collective oscillation the \emph{spin Hall mode}.
In what follows, we investigate this spin Hall mode, focusing
on the time dependence of the spin-dipole moment. In particular, we are
interested in the magnitude of the spin-dipole moment, which builds up
in time.

\section{Analytic results}
\label{sec:analytic}

In this section we present an analytic solution for the ground-state response
to driving when the SOC is either weak or strong compared to the harmonic
trap. Due to time-reversal invariance, all single particle states are doubly
degenerate (Kramer's pairs). Hence, this system has two degenerate many-body
ground states, $|g_1\rangle$ and $|g_2\rangle$, regardless of the strength of
the SOC. Considering different occupation for these two states breaks the
time-reversal symmetry and results in a spurious anomalous Hall effect. We
thus assume that these two ground states have equal occupation.

We follow the unitary evolution of each of the two degenerate 
ground states of the system after 
applying an infinitesimal translation operator
\be
T = 1 - i p_x x_0,
\ee
in the $x$ direction. In general, a convenient method to 
implement unitary time evolution in the Schr\"odinger picture 
is projecting the translated state onto the eigenbasis of the full Hamiltonian $H$. 
To this end, we define the projection operator $\mathcal P_j = | j \rangle \langle j |$ with
respect to the eigenstate $| j \rangle$ of $H$. We are interested in two types of response.
First, we consider of the center-of-mass position. 
For a single state $| s \rangle$, the evolution of the 
average position is given by
\be
\langle x \rangle_s \equiv 
\sum_{j,k}
e^{- i (E_k - E_j) t}
\langle  s | T^\dagger \mathcal P_j^\dagger 
x 
\mathcal P_k T |s \rangle,
\label{eq:exptX}
\ee
and therefore for the two ground states we have
\be
\langle x \rangle =
\frac{1}{2}\sum_{s=g_1,g_2} \langle x \rangle_{s},
\label{eq:avgX}
\ee
where the factor $1/2$ enters as we consider a mixture of both 
ground states with equal probability. 
In the absence of SOC, this expectation value
oscillates with the trap frequency $\omega_T$. 
Second, we investigate the spin-dipole moment, the evolution of
which for a state $| s \rangle$ is
\be
\langle y \sigma_z \rangle_s \equiv
\sum_{j,k}
e^{- i (E_k - E_j) t}
\langle s |  T^\dagger \mathcal P_j^\dagger 
y \sigma_z 
\mathcal P_k T |s \rangle,
\label{eq:exptYSZ}
\ee
and hence for the two ground states we have
\ba
\langle y \sigma_z \rangle
=
\frac{1}{2}\sum_{s = g_1, g_2} \langle y \sigma_z \rangle_s.
\label{eq:avgYSZ}
\ea
The spin-dipole moment is only excited by the
center-of-mass displacement in the presence of SOC.
These two expectation values behave qualitatively differently in the weak and strong SOC regimes.
This is expected since by going from weak to strong SOC, the system undergoes a
dimensional reduction from two-dimensional (2D) to effectively one-dimensional 
(1D) dynamics (see Ref.~\cite{Qi2011} and references therein for more
details).
We consider this behavior in detail below.

\begin{figure}[ht]
\begin{center}
\includegraphics[width=0.49\linewidth]{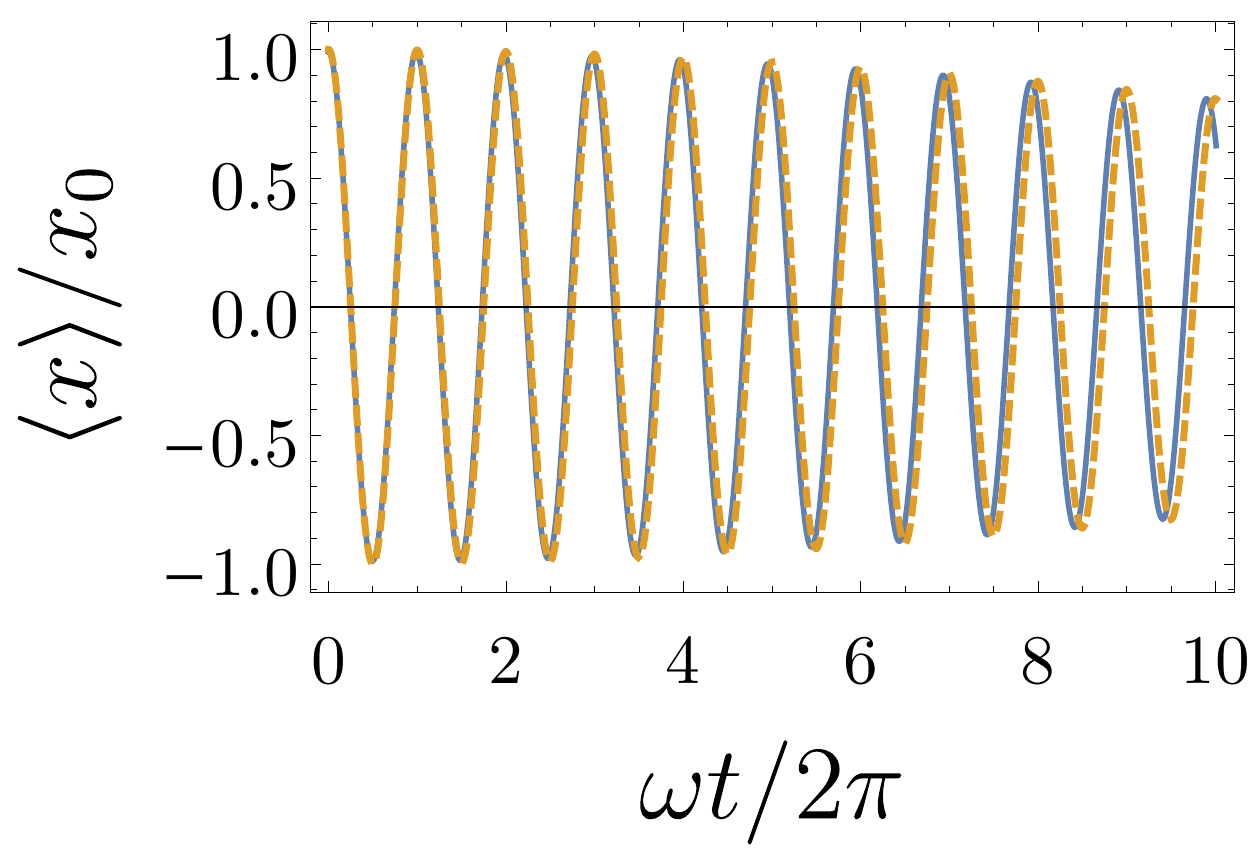}
\includegraphics[width=0.49\linewidth]{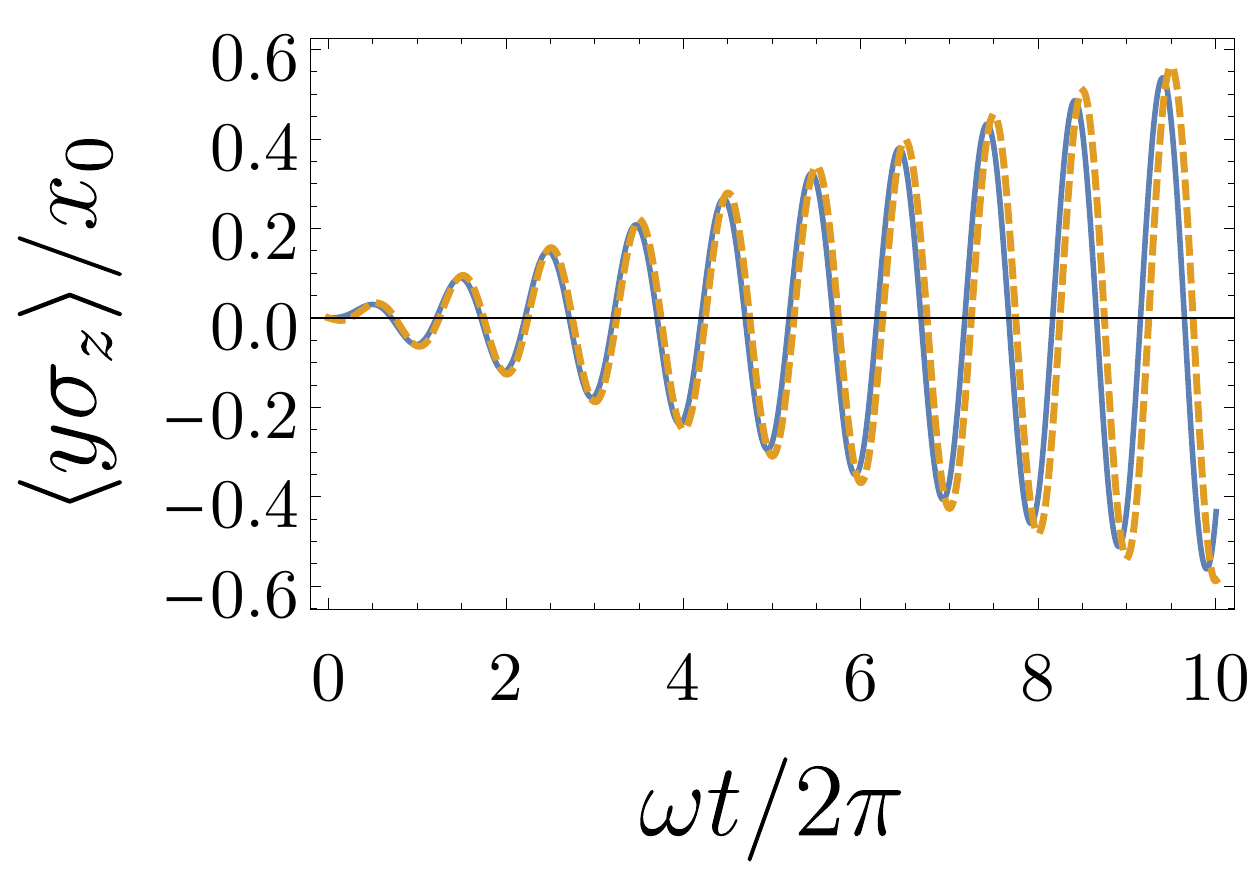}
\caption{
Center-of-mass position and spin-dipole moment oscillations (spin Hall mode)
as a response to displacement of 
the two ground states along the $x$ direction with the displacement
amplitude $x_0$ in the weak SOC regime ($v = 1/10$).
We compare exact numerical results ($x_0 = 1/100$, blue solid line, see Sec.~\ref{sec:numerical}
for more details) 
to the analytical results
in Eqs.~\eqref{eq:weakX}~and~\eqref{eq:weakYSZ} (infinitesimal $x_0$, yellow dashed line) for $10$ trap periods. Fourier transforming these signals and restoring dimensions yields two peaks at 
the energies $\hbar \omega_T \pm Mv^2$.
}
\label{fig:weakComparison}
\end{center}
\end{figure}

\subsection{Weak SOC}
\label{sec:weak}

In the weak SOC limit $v \ll 1$ a natural basis for our problem 
consists of the two-dimensional harmonic oscillator
eigenstates $|n, m, m_s\rangle$, where $n$ is the principal quantum number,
$m$ is the angular momentum quantum number, and $m_s$ is the spin 
projection quantum number.
Nondegenerate perturbation theory can be employed in this case since $m+m_s$ is a
good quantum number \cite{Yin2014}.
Since there are two small parameters, $x_0 \ll 1$ and $v \ll 1$,
capturing the response to the lowest order is especially straightforward in this regime. 
Thus, it is sufficient to consider the second-order energy correction due to $H_R$,
\be
\Delta E_{n, m, m_s} = 
\sum_{n', m', m_s'}^{E_{n,m} \neq E_{n',m'}}
\frac{ \big| \langle n, m, m_s| H_R |n', m', m_s'\rangle \big|^2}
{E_{n,m} - E_{n',m'}}.
\ee
As the translation operator couples the ground state $|0,0,m_s\rangle$ to
the states $|0,\pm 1,m_s\rangle$, the following corrections are required \cite{Yin2014} :
\ba
\Delta E_{0, 0, \uparrow} = -v^2, 
\Delta E_{0, 1, \uparrow} = -2v^2, 
\Delta E_{0, -1, \uparrow} = 0, \\
\Delta E_{0, 0, \downarrow} = -v^2, 
\Delta E_{0, 1, \downarrow} = 0, 
\Delta E_{0, -1, \downarrow} = -2v^2.
\ea
We now take the 2D harmonic oscillator eigenenergies with second-order perturbative 
corrections due to $H_R$ in addition to the unperturbed 2D harmonic 
oscillator eigenstates. Expanding to the lowest nonvanishing correction 
we obtain
\ba
\label{eq:weakX}
\langle x \rangle = x_0 \cos t \cos v^2 t, \\
\label{eq:weakYSZ}
\langle y\sigma_z \rangle = -x_0 \cos t \sin v^2 t,
\ea
in this limit. Hence, taking a Fourier transform of either $\langle x \rangle$ or $\langle y\sigma_z\rangle$ and restoring dimensions yields two peaks at 
the energies $\hbar \omega_T \pm Mv^2$. Note, however, that
in the limit of weak SOC, increasingly long
observation times are required in order to achieve the resolution
sufficient to distinguish the two
peaks in the Fourier spectrum of this signal. Therefore, it might
be more useful to analyze the response directly in the time domain.
We compare these analytical results with a numerical simulation
for $v=1/10$ in Fig.~\ref{fig:weakComparison}. Note that in this weak SOC regime, center-of-mass
oscillations occur around the bottom of the trap with approximately the trap frequency. 
Here SOC introduces 
a modulation with a frequency $v^2$. The spin-dipole response is shifted by a phase of $\pi/2$
with respect to the center-of-mass oscillation. The amplitude of the spin-dipole response
is the same as the amplitude of the center-of-mass oscillation for the two ground states
in this $v \ll 1$ regime. However, note that the spin-dipole moment takes a long time
to build up when $v$ is small, practically limiting the magnitude of the response, see
Fig.~\ref{fig:weakComparison}.
%

\begin{figure}[th]
\begin{center}
\includegraphics[width=0.49\linewidth]{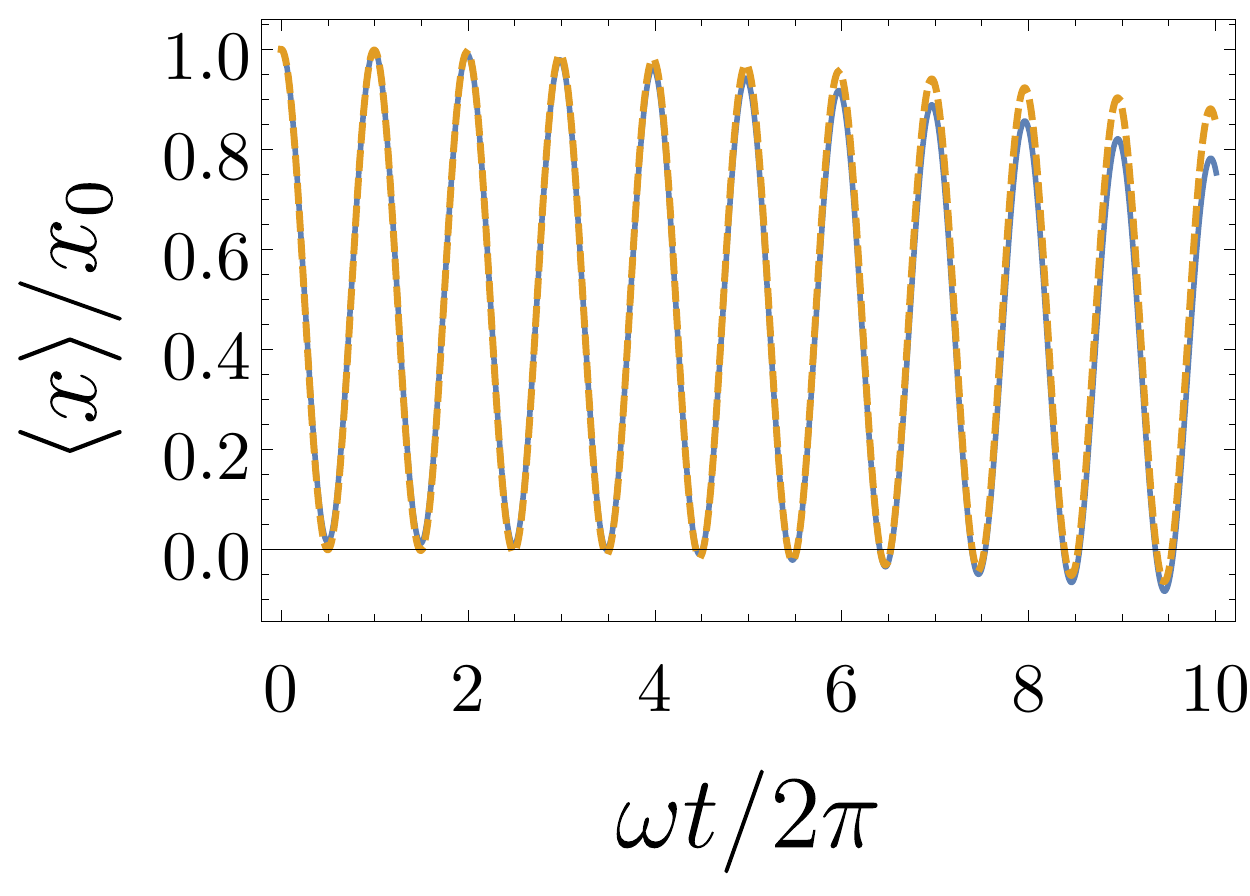}
\includegraphics[width=0.49\linewidth]{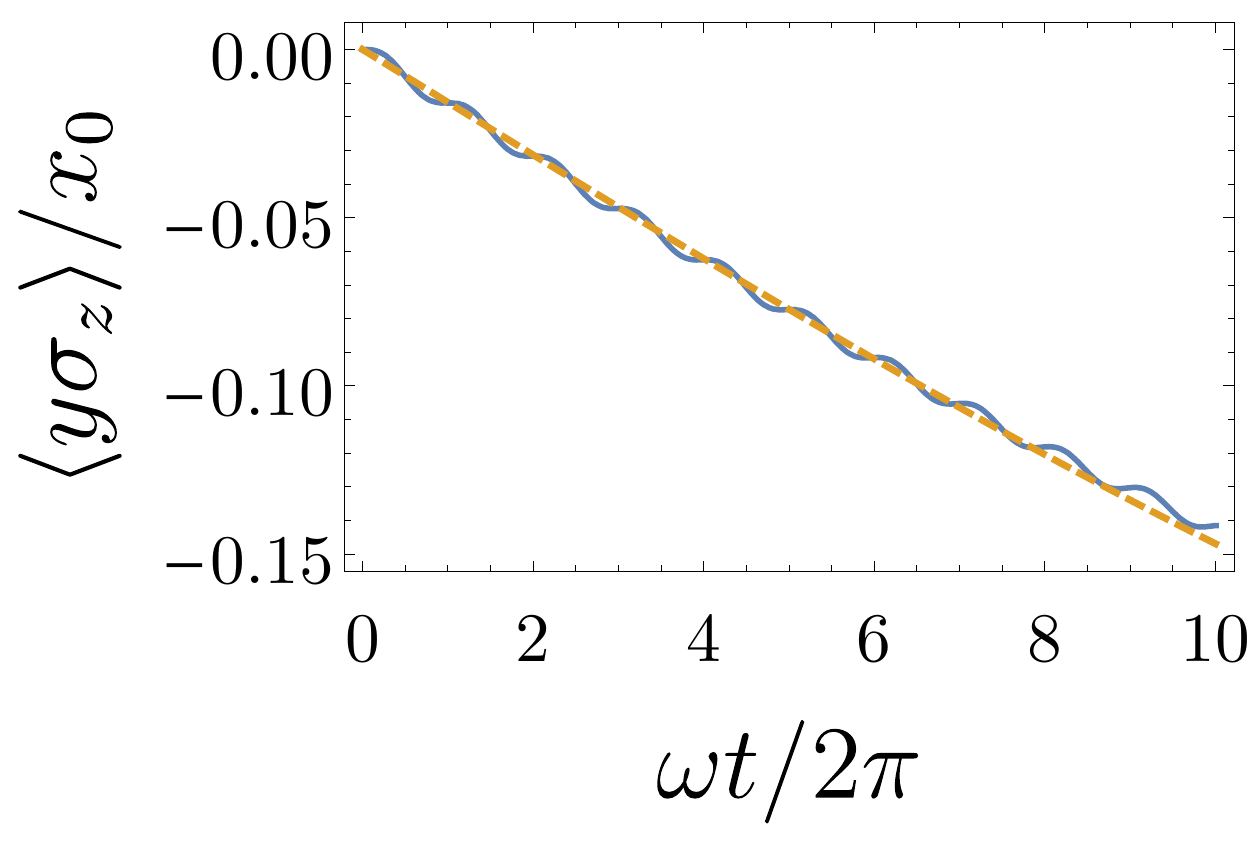}
\caption{
Center-of-mass position and spin-dipole moment oscillations (spin Hall mode)
as a response to displacement of 
the two ground states along the $x$ direction with the displacement
amplitude $x_0$ in the strong SOC regime ($v = 10$).
We compare exact numerical results ($x_0 = 1/100$, blue solid line, see Sec.~\ref{sec:numerical}
for more details) to the analytical results
in Eqs.~\eqref{eq:strongX}~and~\eqref{eq:strongYSZ} (infinitesimal $x_0$, yellow dashed line) for $10$ trap periods.
Fourier transforming the $\langle x \rangle$ signal, discarding
very small frequencies and restoring dimensions yields two peaks at 
the energies $\hbar \omega_T$ and $\hbar \omega_T + 1/Mv^2$.
}
\label{fig:strongComparison}
\end{center}
\end{figure}

\subsection{Strong SOC}
\label{sec:strong}

In this section we apply the procedure of dimensional reduction
pioneered in the field of topological insulators \cite{Qi2011} to our system. 
It allows us to obtain
the spectrum and the wavefunctions explicitly
in the strong SOC regime $1/v\ll 1$. The spectrum of the system in this
limit has already been presented in, e.g., Ref.~\cite{Sinha2011}.

In preparation to treating the problem in the strong SOC regime, it
is convenient to write down the Schr\"odinger equation in 
momentum space,
\ba
\Big(\frac{p^{2}}{2} 
+ ivp 
& 
\left(\sigma_{+}e^{-i\theta}-\sigma_{-}e^{i\theta}\right)
\nonumber \\
&
-\frac{1}{2}
\left(\frac{1}{p}\partial_{p}\left[p\partial_{p}\right]\right)-\frac{1}{2}\frac{1}{p^{2}}\partial_{\theta}^{2}
\Big)\Psi
=E\Psi.
\label{eq:se}
\ea
We now make an Ansatz
\be
\Psi=\psi_{+}(p)\chi_{+}(\theta)e^{-im\theta}+\psi_{-}(p)\chi_{-}(\theta)e^{-im\theta},
\ee
which allows us to see that different $m$ states are not coupled
since
\ba
-\partial_{\theta}^{2}\left(\chi_{\pm}e^{-im\theta}\right)=
\left(m\left[m+1\right]+\frac{1}{2}\right)\chi_{\pm}e^{-im\theta}
\nonumber \\
-\left(m+\frac{1}{2}\right)e^{-im\theta}\chi_{\mp}e^{-im\theta},
\ea
and
\ba
\left[\frac{p^{2}}{2}+ivp\left(\sigma_{+}e^{-i\theta}-\sigma_{-}e^{i\theta}\right)\right]\Psi
=E_{+}\psi_{+}\chi_{+}e^{-im\theta}
\nonumber \\
+E_{-}\psi_{-}\chi_{-}e^{-im\theta}.
\ea
Projecting Eq.~\eqref{eq:se} to the two branches yields
\ba
E_\pm \psi_\pm
-
&
\frac{1}{2}
\left(\frac{1}{p}\partial_{p}\left[p\partial_{p}
\psi_\pm
\right]\right)
\nonumber \\
+
\frac{1}{2p^{2}}
&
\left[\left(m\left[m+1\right]+\frac{1}{2}\right)\psi_\pm-\left(m+\frac{1}{2}\right)\psi_\mp\right]
=E\psi_\pm,
\ea
where $E_\pm$ was defined in Eq.~\eqref{eq:rashbaDisp}.
Note that all the considerations up to this point have been exact. Now we specialize
to the strong SOC regime, where the two branches are separated by
a large energy gap, save for the $p=0$ point. Relying on this fact,
we assume that the upper branch is empty, $\psi_{+}=0$, while the
lower branch is described by the wavefunction
\be
\psi_{-}=f(p)/\sqrt{p}.
\ee
The Schr\"odinger equation for $f(p)$ thus is
\be
E_{-}f-\frac{f}{8p^{2}}-\frac{1}{2}\partial_{p}^{2}f+\frac{1}{2p^{2}}\left(m\left[m+1\right]+\frac{1}{2}\right)f=Ef.
\ee
In order to investigate the low-lying states close to the Rashba ring
$p=v$, we complete the square,
\ba
-\frac{1}{2}\partial_{p}^{2}f+\left(\frac{p^{2}}{2}-pv+\frac{v^{2}}{2}\right)f-\frac{f}{8v^{2}}-\frac{v^{2}}{2}f
\nonumber \\
+\frac{1}{2p^{2}}\left(m\left[m+1\right]+\frac{1}{2}\right)f=Ef,
\ea
and notice that the first two terms on the left-hand side describe
a one-dimensional harmonic oscillator. For the remaining terms, we
approximate $p\simeq v$ and obtain the following spectrum:
\be
E_{\nu m}=\left(\nu+\frac{1}{2}\right)-\frac{v^{2}}{2}+\frac{1}{2v^{2}}\left(m\left[m+1\right]+\frac{1}{4}\right),
\label{eq:strongSOCspectrum}
\ee
where $\nu$ is the quantum number of the 1D harmonic oscillator. Note that this spectrum
preserves the degeneracy of the Kramer's pairs and in particular there
are two ground states with the energy, 
\be
E_{00}=E_{0-1}=\frac{1}{2}-\frac{v^{2}}{2}+\frac{1}{8v^{2}}.
\ee
This spectrum has been shown to match the strong SOC spectrum for
the low-energy states very well \cite{Sinha2011, Hu2012b}. In this approximation,
momentum-state wavefunctions are
\be
\Psi_{\nu m}=f_{\nu}(p-v)\frac{1}{\sqrt{2p}}\begin{pmatrix}-ie^{-i\theta}\\
1
\end{pmatrix}\frac{e^{-im\theta}}{\sqrt{2\pi}},
\ee
where
\be
f_{\nu}(p)=\frac{1}{\sqrt{2^{\nu}\nu!}}\frac{1}{\pi^{1/4}}e^{-p^{2}/2}H_{\nu}(p)
\ee
are the 1D harmonic oscillator eigenstates, and $H_{\nu}$ denotes the Hermite polynomial. 

We now apply the translation operator in the $x$ direction on the two ground states,
\ba
T\Psi_{00} & =  \Psi_{00} 
\nonumber \\ & - \frac{ix_{0}}{2} \left(\frac{\Psi_{11}}{\sqrt{2}}+\frac{\Psi_{1-1}}{\sqrt{2}}+v\Psi_{01}+v\Psi_{0-1}\right),
\ea
\ba
T\Psi_{0-1} & = \Psi_{0-1}
\nonumber \\
& -\frac{ix_{0}}{2}
\left(
\frac{\Psi_{10}}{\sqrt{2}} + \frac{\Psi_{1-2}}{\sqrt{2}}
+ v\Psi_{00} + v\Psi_{0-2}
\right),
\ea
where we have used one of the Hermite function recursion relations,
namely,
\be
pf_{n}(p)=\sqrt{\frac{n}{2}}f_{n-1}(p)+\sqrt{\frac{n+1}{2}}f_{n+1}(p).
\ee
Subsequently, Eqs.~\eqref{eq:avgX}~and~\eqref{eq:avgYSZ} yield
\ba
\label{eq:strongX}
\left\langle x\right\rangle &= \frac{x_{0}}{4}\left(\cos t+2\cos\frac{t}{v^{2}}+\cos\left[t+\frac{t}{v^{2}}\right]\right),
\\
\left\langle y\sigma_{z}\right\rangle & = -\frac{x_{0}}{4}\sin\frac{t}{v^{2}}.
\label{eq:strongYSZ}
\ea
Taking a Fourier transform of the $\langle x \rangle$ signal, discarding
very small frequencies and restoring dimensions yields two peaks at 
the energies $\hbar \omega_T$ and $\hbar \omega_T + 1/Mv^2$.
We compare these analytical results with a numerical simulation
for $v=10$ in Fig.~\ref{fig:strongComparison}. Note that in this strong SOC regime, center-of-mass
oscillations are qualitatively different from the dipole mode in the absence of SOC. In particular,
here the center-of-mass position oscillates around the initial position $x_0$ and not the bottom of 
the trap \cite{Zhang2012a} which is ultimately due to the physical momentum
being substantially different from the canonical momentum 
\cite{Armaitis2017b}. Note further that the amplitude of the spin-dipole response is only one quarter of that of 
center-of-mass oscillations.

\section{Exact numerical results}
\label{sec:numerical}

While at zero temperature for small perturbation amplitude $x_0\ll 1$ 
and extreme SOC strengths $v \ll 1$ and $v \gg 1$ we have managed to obtain 
analytical results, other regimes remain unexplored. 
To address this, we turn to numerically evaluating              
the evolution of the system. In this way, we can cover                 
perturbation amplitudes and SOC strengths of any magnitude.
We consider, moreover, nonzero temperature. 

We now investigate the amplitude of the spin-dipole moment oscillations in
response to a small (one tenth of the trap length) displacement of the
center-of-mass position of a system of either bosons or fermions.  We limit
the time of evolution to ten trap periods so as to account for the
finite lifetime of ultracold atomic samples in the experiment.  In order to
explore qualitative effects of statistics and temperature on the spin Hall
mode, we consider a system of $100$ particles at three temperatures, $k_B T =
1/10, 1, 3/2$ in the units of trap energy, where $k_B$ is the Boltzmann
constant. The number of particles is limited by computational requirements. 

In practice, we fix the number of particles $N = 100$ and
solve
\be
N = \sum_s 
\frac{1}{\exp\left( \left[E_s - \mu \right] / k_B T \right) \mp 1  }
,
\ee
for the chemical potential $\mu$,
where the sum runs over the eigenstate energies $E_s$, and 
the upper (lower) sign corresponds to bosonic (fermionic)
statistics. Subsequently, we compute how the
expectation values evolve in time:
\ba
\langle x \rangle = \frac{1}{N} 
\sum_s 
\frac{\langle x \rangle_s}{\exp\left( \left[E_s - \mu \right] / k_B T \right) \mp 1  },
\\
\langle y\sigma_z \rangle = \frac{1}{N} 
\sum_s 
\frac{\langle y\sigma_z \rangle_s}
{\exp\left( \left[E_s - \mu \right] / k_B T \right) \mp 1  },
\ea
where the time-dependent center-of-mass position and
spin-dipole moment for a state $s$, namely,
$\langle x \rangle_s$ and $\langle y\sigma_z \rangle_s$,
have been defined in Eqs.~\eqref{eq:exptX}~and~\eqref{eq:exptYSZ}.
We emphasize that these quantities are evaluated exactly in our 
non-interacting system \cite{zaltys}.
Since the results for bosons and fermions are quite different,
we present them one after another.

\begin{figure}[ht]
\begin{center}
\includegraphics[width=0.99\linewidth]{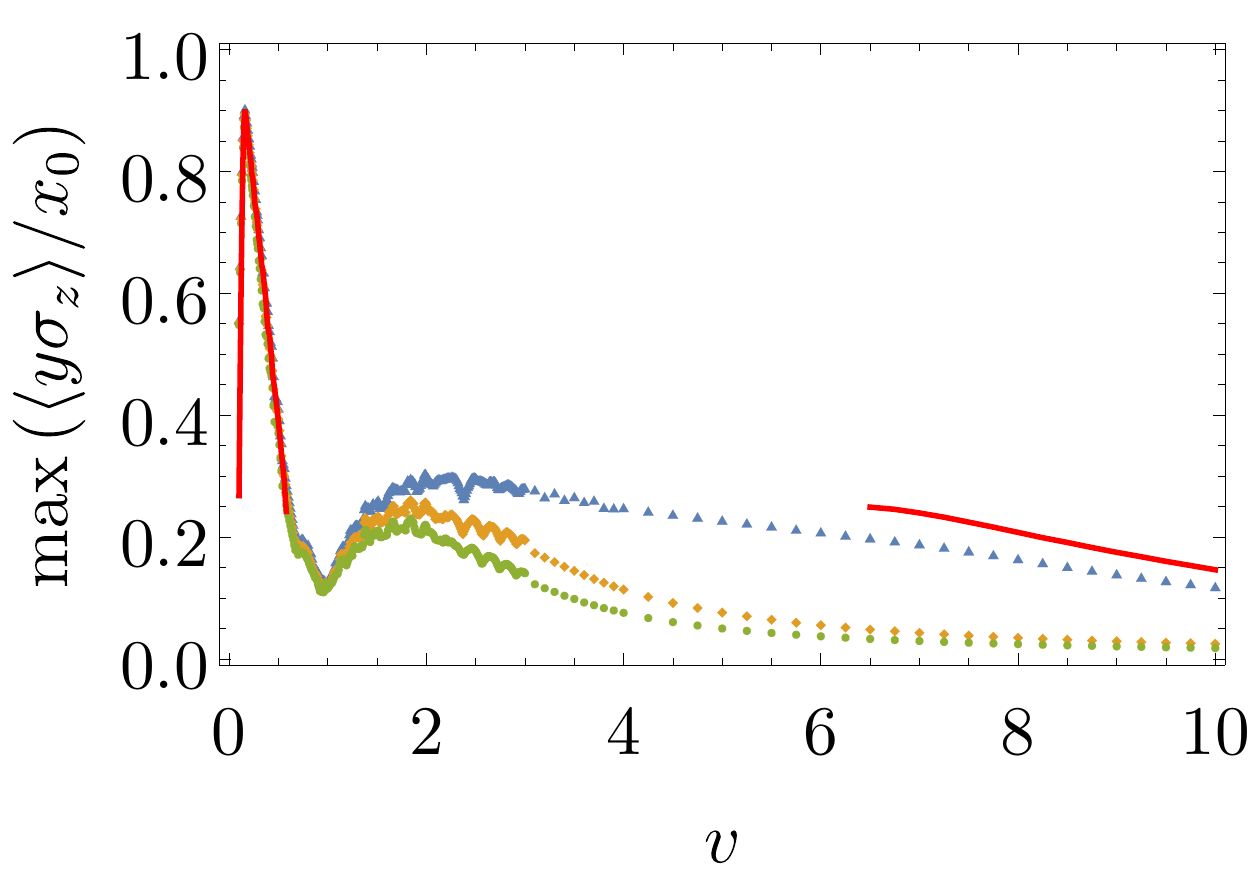}
\caption{
Maximum amplitude of the transverse spin-dipole moment 
$\langle y \sigma_z \rangle$ for $100$ bosons, normalized to the initial
center-of-mass displacement amplitude $x_0$ (equal to $1/10$ of the trap 
length),
observed over the time interval of ten trap periods. Three temperatures
in the units of the trap energy have been investigated: 
$k_B T = 0.1$ (blue triangles), $1$ (orange diamonds), and $1.5$ (green
circles). At weak and strong SOC, the response can be described
analytically at zero temperature
(red solid curves, Eqs.~\eqref{eq:weakSOCansatz}~and~\eqref{eq:strongYSZ},
see Sec.~\ref{sec:numerical} for more details).
}
\label{fig:boseThermal}
\end{center}
\end{figure}

\begin{figure}[ht]
\begin{center}
\includegraphics[width=0.49\linewidth]{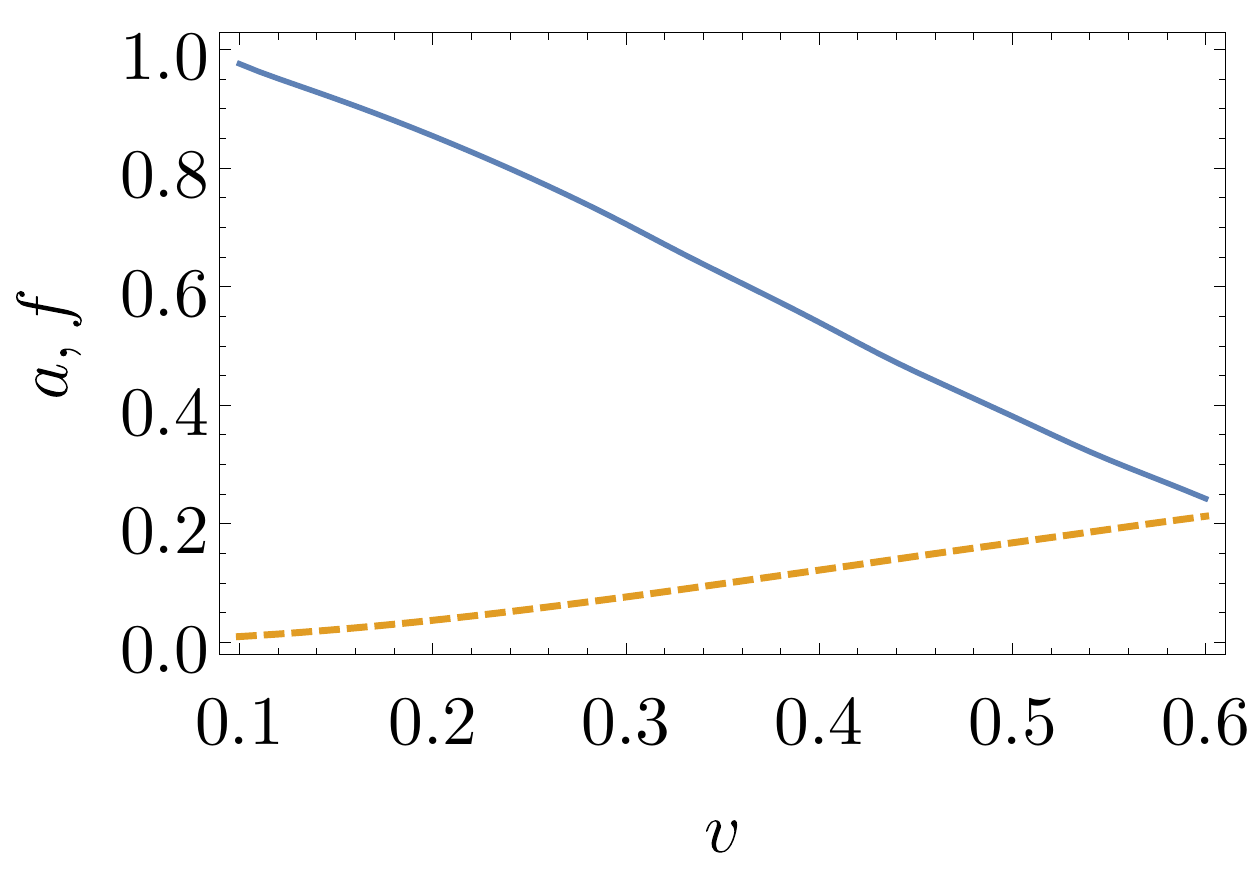}
\includegraphics[width=0.49\linewidth]{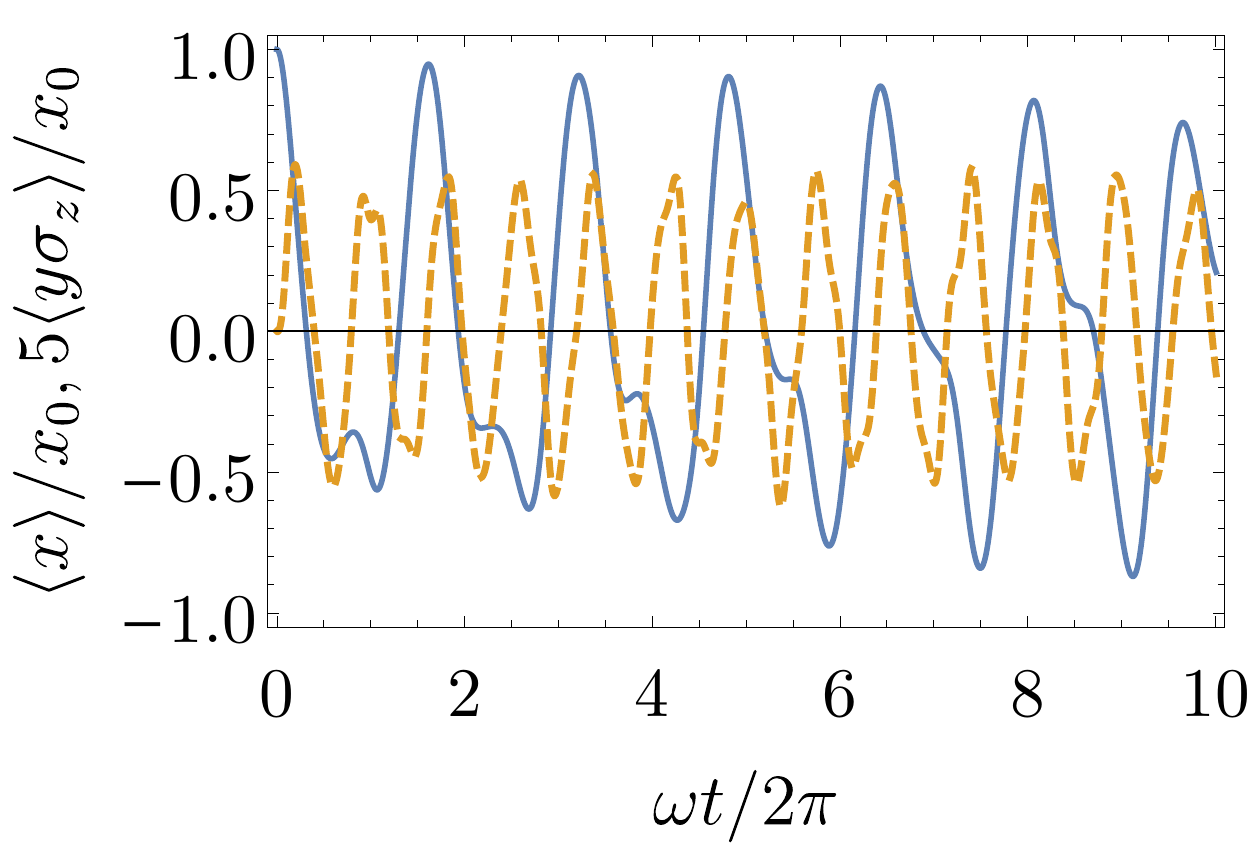}
\caption{
Left: the amplitude $a$ (blue solid line) and 
the frequency $f$ (yellow dashed line) fits 
for the spin-dipole moment Ansatz in Eq.~\eqref{eq:weakSOCansatz}
of $100$ bosons.
Right: the center-of-mass position (blue solid line) and the 
transverse spin-dipole moment
(yellow dashed line; multiplied by a factor of $5$ for clarity) 
for $100$ bosons at $v=1$ (intermediate SOC strength), where the 
amplitude of the spin-dipole moment response shows a local minimum for bosonic particles at all considered temperatures.
}
\label{fig:boseFitMinimum}
\end{center}
\end{figure}

For bosons the amplitude of the response is given in 
Fig.~\ref{fig:boseThermal}. For all investigated temperatures, the
plot can be roughly divided into three parts: weak, strong and
intermediate SOC strength. Whereas the explanation of the
first two regimes are relatively straightforward, the same cannot be said
about the intermediate regime of $0.6 \lesssim v \lesssim 7$.
In the limit of weak SOC (small $v$), the evolution of the 
spin-dipole moment displays a very weak temperature dependence and
can be described analytically (Sec.~\ref{sec:weak}). This 
approximation is only accurate up to the first peak, which occurs at
$v \simeq 0.15$. However, by employing Eq.~\eqref{eq:weakYSZ} as an Ansatz we are able
to fit the response up to $v \simeq 0.6$. Explicitly, we use
\be
\langle y\sigma_z \rangle = -a(v)\, \cos t \sin f(v)\, t,
\label{eq:weakSOCansatz}
\ee
where now the amplitude $a$ and the frequency $f$ depend 
on the SOC strength,
see Fig.~\ref{fig:boseFitMinimum} (left). For small $v$, the response is 
virtually independent
of temperature in the range that we have investigated, as almost 
exclusively the two degenerate ground states are occupied. 
This is because the spectrum
in this regime consists of weakly-perturbed harmonic oscillator states,
therefore resulting in exponentially suppressed occupation of excited
states for bosons.
We are also able to explain the results in the limit of strong SOC $v
\gtrsim 7$ at low temperature, as there the spin-dipole moment
approximately follows the evolution described in Eq.~\eqref{eq:strongYSZ}.
However, in this regime the match is not perfect at any nonzero 
temperature,
and it worsens as the system is heated. This is because the gaps
in the spectrum decrease for moderate SOC strength
due to effective dimensional reduction [cf.~Eq.~
\ref{eq:strongSOCspectrum}], and thus excited states are readily 
occupied. Note that this is also the reason why the effect
of temperature on the response is the strongest for moderately
strong SOC.
When several states are occupied, the behavior becomes more complicated.
In particular, since different-energy states contribute oscillations of
different frequencies and phases to the response, it is not straightforward
to explain the minimum and the maximum shown in Fig.~\ref{fig:boseThermal}
for the intermediate values of $v$. To illustrate the typical response
in this intermediate regime, in Fig.~\ref{fig:boseFitMinimum} (right) we show the evolution of 
the transverse spin-dipole moment for $v=1$, which is the local response minimum for all 
considered temperatures.

\begin{figure}[ht]
\begin{center}
\includegraphics[width=0.99\linewidth]{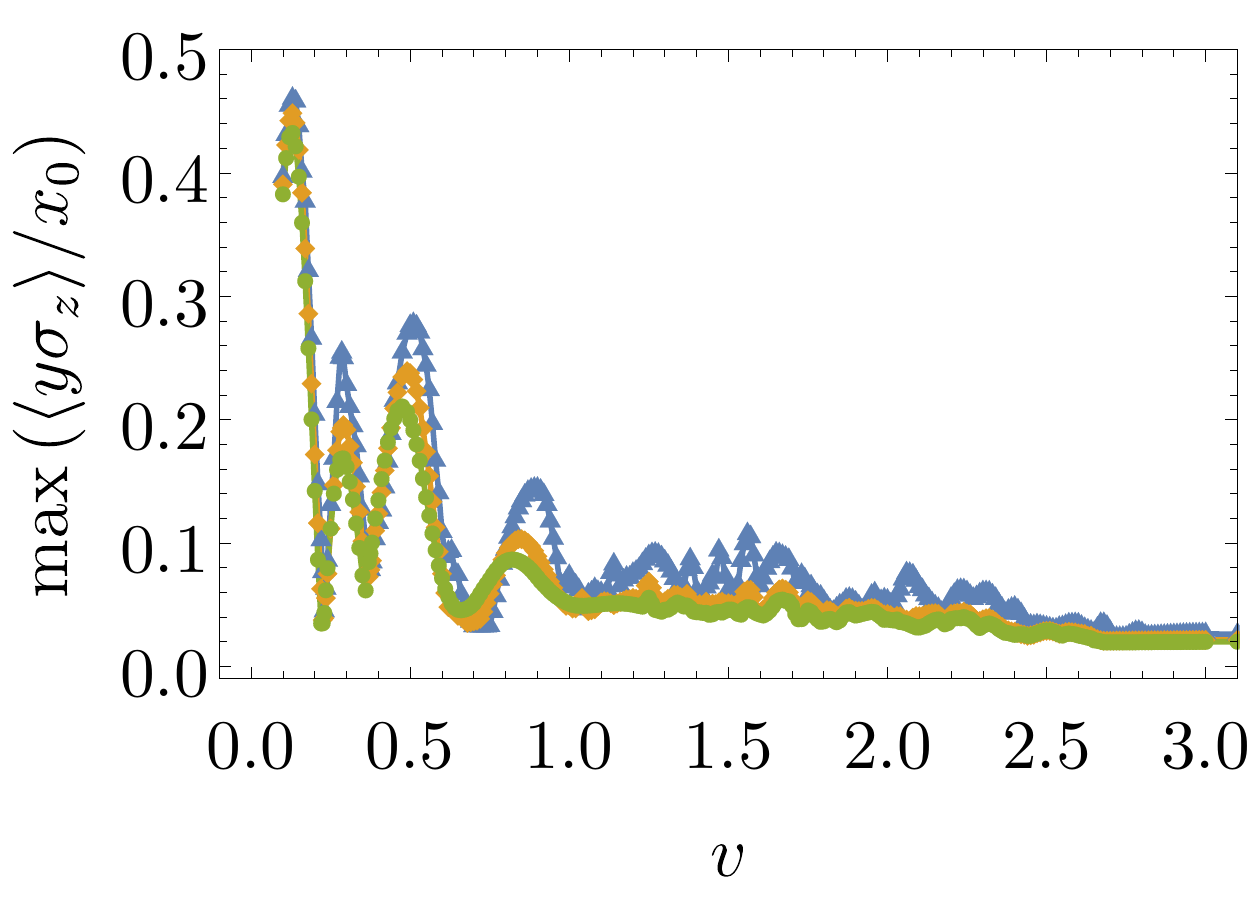}
\caption{
Maximum amplitude of the transverse spin-dipole moment 
$\langle y \sigma_z \rangle$ for $100$ fermions, normalized to the initial
center-of-mass displacement amplitude $x_0$ (equal to $1/10$ of the trap 
length),
observed over the time interval of ten trap periods. Three temperatures
in the units of the trap energy have been investigated: 
$k_B T = 0.1$ (blue triangles), $1$ (orange diamonds), and $1.5$ (green circles). The points at each temperature
are joined by a line of the corresponding color for clarity.
}
\label{fig:fermiThermal}
\end{center}
\end{figure}

Occupation of several states is particularly relevant for fermions
(Fig.~\ref{fig:fermiThermal}). In that case, the interference
of responses of various states plays a crucial role, resulting
in several local minima and maxima, which become
sharper as temperature decreases. However, the qualitative
structure of the response is similar to bosons, in the sense
that there is a global maximum at $v \simeq 0.13$ (compare
to $v \simeq 0.15$ for bosons), and the response becomes
progressively weaker for stronger SOC.
In the region $3 < v < 10$ the response is below $0.05$
and decays very slowly with increasing SOC strength
for all considered temperatures, and hence we have excluded
this region from the plot.

\begin{figure}[ht]
\begin{center}
\includegraphics[width=0.99\linewidth]{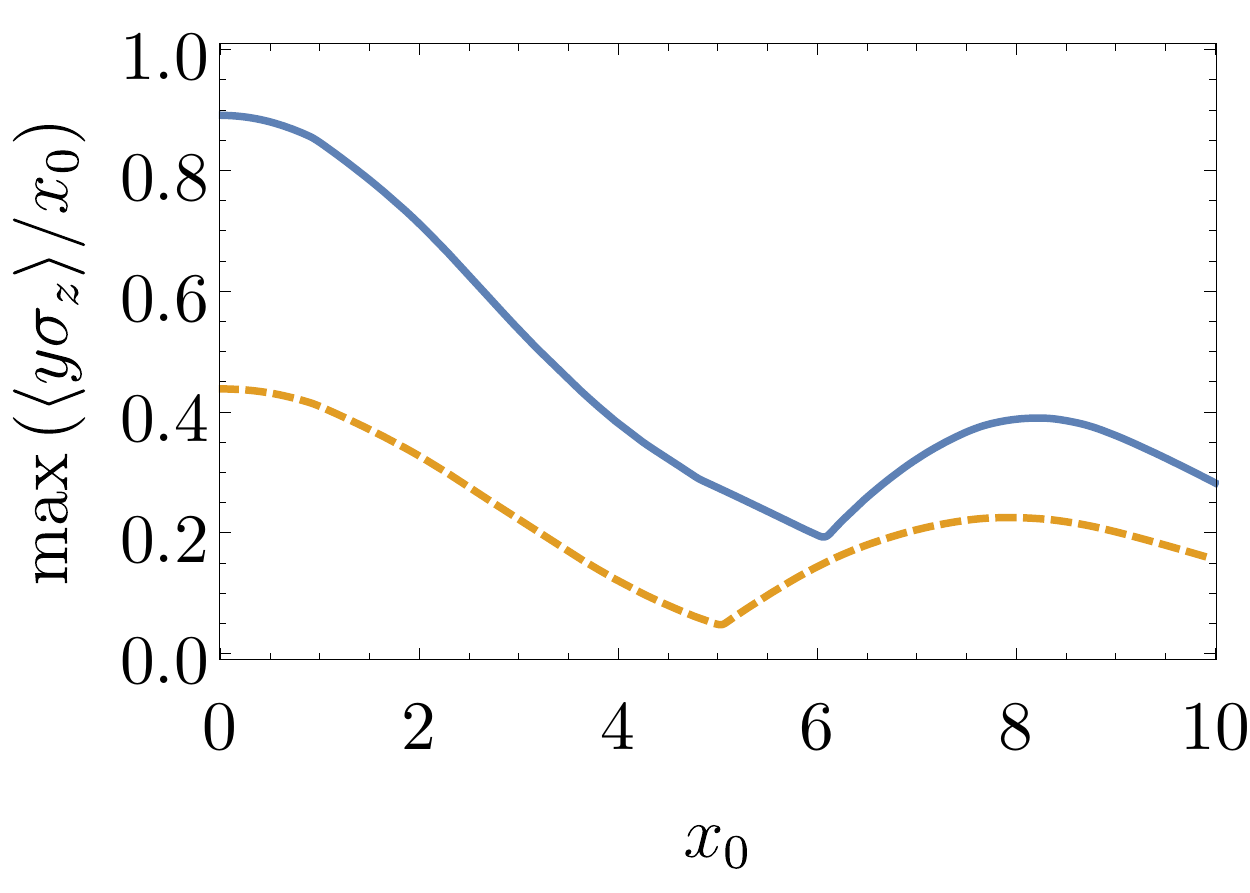}
\caption{
Maximum normalized amplitude of the transverse spin-dipole moment 
$\langle y \sigma_z \rangle$ for $100$ bosons (solid blue line)
and fermions (dashed yellow line) at a temperature
of $k_B T = 1/10$ and SOC strength $v = 0.15$ observed over the time interval 
of ten trap periods
as a function of the initial
center-of-mass displacement amplitude $x_0$.
}
\label{fig:nonlinear}
\end{center}
\end{figure}

Finally, we would like to emphasize that all the results
reported thus far are in the limit of small initial displacement.
The response of the system is nonlinear in the sense that
the maximum amplitude of the spin-dipole moment oscillation strongly
depends on the initial displacement, Fig.~\ref{fig:nonlinear}.

\section{Summary and outlook}
\label{sec:summary}

We have investigated a harmonically-trapped system with Rashba SOC
and no interparticle interactions. In particular, we have studied
the response of the system to a small displacement away from the bottom
of the trap. We found that in addition to the expected center-of-mass
oscillations, a dynamics of the spin-dipole moment is induced.
This spin-dipole moment dynamics is transverse to the displacement
direction, and is analogous to the spin Hall effect.
Therefore, we dubbed this collective mode the spin Hall mode.

Furthermore, we have performed an exact numerical study of
the qualitative effects of temperature and
statistics on the amplitude of the spin-dipole moment oscillations. 
For bosons in the weak-SOC and strong-SOC limits, the response
is captured by analytic expressions. In the intermediate-SOC region
for bosons, as well as for fermions at any SOC strength, 
the spin-dipole moment oscillations appear anharmonic. Even though the 
amplitude of these oscillations 
as a function of the SOC strength is different for bosons
and fermions, we have found that
in regions where the spin Hall response is the strongest,
the effects of temperature and statistics are weak. We hope that
our analysis will stimulate experimental work on
collective modes of ultracold atomic gases with 2D SOC.

In future work, building on more formal results 
\cite{Tutunculer2004}, it might be
possible to extend the simple analytical treatment presented here. 
Furthermore, as some of the realistic Rashba SOC implementation schemes
\cite{Campbell2011} might result in various anisotropies, it is both
feasible \cite{Marchukov2013,Marchukov2014,Marchukov2015} 
and desirable to investigate the effects of such anisotropies on the
spin Hall mode in an approach similar to the one presented here.
In order to make quantitative predictions for experiments,
in addition to anisotropies, careful accounting for the nonlinear
behavior of the system is important.
Exploring interaction effects, for example, in a mean-field
(Hartree-Fock) type of treatment is another promising research direction.
Finally, one could investigate how the results presented here are altered
in a Bose-Einstein-condensed phase or in the presence
of pair condensation.

\acknowledgements

It is our pleasure to thank
Dimitrios Trypogeorgos,
Ian Spielman,
Lev Pitaevskii,
Oleksandr Marchukov,
and Sandro Stringari
for stimulating discussions.
J.~A. has received funding from the European Union's Horizon 2020 research and innovation
programme under the Marie Sk\l odowska-Curie grant agreement No 706839 (SPINSOCS).
R.~A.~D. and H.~T.~C.~S. are supported by the Stichting voor Fundamenteel Onderzoek der
Materie (FOM) and are part of the D-ITP
consortium, a program of the Netherlands Organisation for Scientific Research (NWO)
that is funded by the Dutch Ministry of Education, Culture and Science (OCW).
R.~A.~D. is also supported by the European Research Council (ERC).

\bibliography{spinhall-mode}

\end{document}